\documentclass[a4paper]{aa}
\usepackage{aabib, epsfig}

\begin{document}
\thesaurus{03.09.1; 08.16.7 Crab} 

\title{Optical STJ Observations of the Crab Pulsar}

\author{M.A.C.~Perryman, F.~Favata, A.~Peacock, N.~Rando, B.G.~Taylor}

\institute{Astrophysics Division, Space Science Department of ESA, ESTEC,
  Postbus 299, NL-2200 AG Noordwijk, The Netherlands
}


\date{Received date 19 April 1999; accepted date 1 May 1999}

\titlerunning{Crab Pulsar observed with STJs}
\authorrunning{M.A.C.$\,$Perryman et al.}

\maketitle 
\begin{abstract}
We report the first observations of an astronomical object using a
superconducting tunnel junction (STJ) device, a pixel detector with
intrinsic energy resolution in the optical wavelength range. The Crab
pulsar was observed using a $6\times6$ array of Tantalum STJs at the
4.2-m William Herschel Telescope on La Palma. Each array element
provides photon counting capability recording up to $\sim10^3$~photons
pix$^{-1}$ s$^{-1}$ with an arrival time accuracy of about 5~$\mu$s,
and providing a wavelength resolution of about 100~nm. We derive a
photometrically resolved light curve which, however, shows no
significant colour index variations with pulsar phase.

\end{abstract}

\section{Introduction}
\label{sec:intro}

The possibility of intrinsic determination of individual photon
energies in the optical range was first reported by \cite*{pfp93}, who
proposed the application of STJ technology to optical photon counting.
Incident photons break Cooper pairs responsible for the
superconducting state. Since the energy gap between the ground state
and excited state is only a few meV (rather than $\sim1$~eV in the
case of semiconductors), each individual photon creates a large number
of free electrons, in proportion to the photon energy. The first
experiments demonstrating single optical photon counting with energy
resolution were reported by \cite*{pvr+96} using STJs, and
further developments have been described by \cite*{pvr+97}. Similar
results using superconducting transition-edge sensors (TES) as
microcalorimeters have recently been reported (\cite{ccc+98}), 
including first observations of the Crab pulsar by \cite*{rom+98}.

Technology development within the Astrophysics Division at ESA is
ultimately aiming at large arrays of low $T_{\rm c}$ superconductors
capable of $\sim10$~\AA\ intrinsic energy resolution at high count
rates. A $6\times6$ array of $25\times25$~$\mu$m$^2$ Tantalum
junctions has been developed as a first astronomical prototype
(\cite{rpa+98}). The wavelength response is intrinsically very broad
(from $<300$~nm to $>1000$~nm) but is restricted in the present system
to about 300--700~nm, as a result of the atmosphere ($\sim300$~nm) and
the optical elements required for the suppression of infrared photons
($\sim700$~nm). The detector quantum efficiency is around 70\% across
this wavelength range, limited by the device/substrate geometry rather
than by the intrinsic detector response. Count rate limits are about
$10^3$~photons~s$^{-1}$, determined by the output stage electronics,
although the device relaxation time is much faster, being below
$\sim10~\mu$s for the present device. The current wavelength
resolution, $\sim100$~nm at 500~nm, is driven by system electronics
and residual thermal background (IR) radiation, although the intrinsic
response of the Ta~junctions is some factor of 5~better than the
present performance.

PSR~B0531+21 in the Crab Nebula was first observed as an optical pulsar 
by \cite*{cdt69}, and provides an excellent target for verification of 
the system's astronomical performance. Along with PSR~B0833--45 in
Vela (\cite{wal+77}), PSR~B0540--69 in the LMC (\cite{mp85}) and more
recently PSR~B0656+14 (\cite{srg+97}) and possibly Geminga
(\cite{sgh+98}) it remains one of the few pulsars observed to emit
pulsed optical radiation.

While the $\sim33$~ms period pulsar has been extensively studied at all
wavelengths including the optical (e.g.\ \cite{pbd+93}, \cite{efr+96}, 
\cite{nmc+96}, \cite{ef97}, \cite{glc+98}, \cite{mhs98}) it continues to 
be important in providing new insights into the nature of the pulsar
emission mechanism: the pulse profile shape, the separation of the
primary and secondary emission peaks by 0.4~in phase and recent results
on the energy dependence of the pulse shape over the infrared to 
ultraviolet range (\cite{efr+96}, \cite{ef97}) provide a challenge to 
theoretical models in which $\gamma$-rays created through curvature 
radiation interact with the pulsar magnetosphere to produce the
X-ray, ultraviolet, optical and infrared pulsations through a variety 
of energy-loss mechanisms. A photon counting detector with
energy resolution in the optical offers an important possibility to
examine further the energy dependence as a function of pulse phase.

\section{Observations}
\label{sec:observations}

Our prototype $6\times6$ Tantalum STJ array covering an area of
$\sim4\times4$~arcsec$^2$ was operated at the Nasmyth focus of the
William Herschel Telescope on La Palma in February 1999. Photon
arrival time information was recorded with an accuracy of about
$\pm5$~$\mu$s with respect to GPS timing signals; while the latter is
specified to remain within 1~$\mu$s of UTC, typical standard deviations
are much less (\cite{kus96}). Observations were made on 4--6~Feb,
although modest seeing ($>2$~arcsec), especially poor on the first two
nights, and a significant number of unstable junctions, meant that
total intensities could not be determined reliably. Our present
analysis is restricted to the signal extracted from just 6~pixels
(corresponding to an indeterminate but small ($\sim0.1$) fraction of
the overall PSF), and a consideration of the resulting pulse profile
and its energy dependence.

Data from the STJ are archived in FITS format with the photon records
defined by their arrival time, $x,y$ pixel coordinate, and 
energy channel in the range 0--255. Channels 50--100 cover 
$\Delta\lambda\sim610-310$~nm, with 
$\lambda({\rm nm})\sim1238.5/(m\times N_{\rm ch}+c)$,
where $N_{\rm ch}$ is the channel number, $m\sim0.04$, and 
$c\sim0.03$. Energy calibration was performed using an internal 
calibration source before and after the target observations, and 
verified using narrow-band filter observations of a standard star.

Photon arrival times were translated to the
solar system barycentre using the JPL DE200 ephemeris, taking into
account gravitational propagation delay. Our reference timing
ephemeris for the Crab pulsar used the 15~Feb 1999 (MJD~=~51224) 
values of $\nu=29.856\,514\,436\,4$~Hz and 
$\dot\nu=-374\,886.90\times10^{-15}$~s$^{-2}$ taken from the radio
ephemeris of \cite*{lpr99}. Consistent periods were obtained, with a
precision of typically $5\times10^{-8}$~s, from a period search of 
the timing data. 

\begin{figure}[thbp]
\begin{center}
\leavevmode 
\epsfig{file=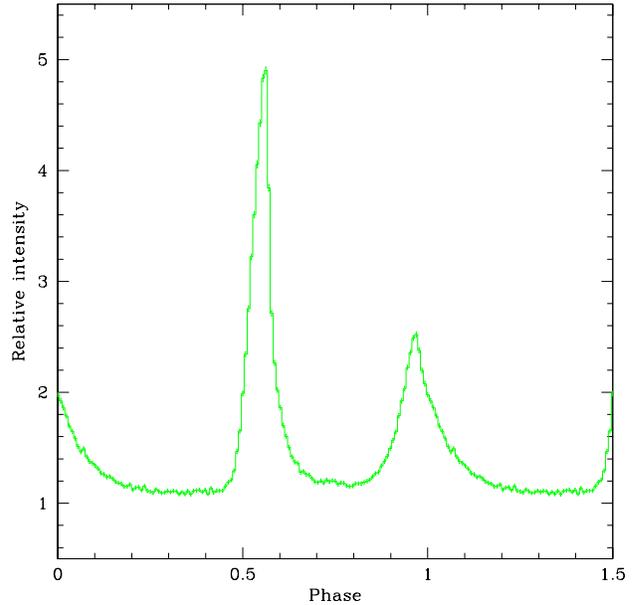, width=8.8cm}
\caption{Pulse profile for the Crab pulsar from the 6~Feb 1999 data 
over the range 310--610~mn (128 phase bins). Photon statistical error bars
are included. During the 10~min interval of best seeing, the total 
counts summed over the 6~active pixels corresponds to about 1000 
photons s$^{-1}$, roughly half of which originate from the unpulsed
component (including sky) and half from the pulsed component.
} \label{fig:lightcurve}
\end{center}
\end{figure}

\begin{figure}[bhtp]
\begin{center}
\leavevmode 
\epsfig{file=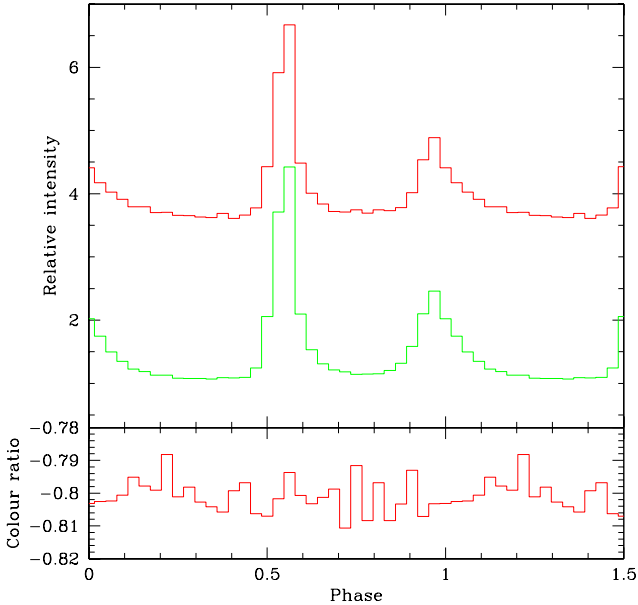, width=8.8cm}
\caption{Normalised 32-phase bin pulse profiles for the 6~Feb data 
(i.e.\ energy cuts of Fig.~1), divided into the E$_1$~=~310--410~nm
(upper curve) and E$_2$~=~500--610~nm (middle curve) energy channels,
with $\lambda_{\rm eff}\sim378$ and 553~nm respectively. The upper
curve has been shifted vertically by 1.0. The resulting colour ratio versus 
phase, constructed as E$_2$--E$_1$/E$_2$+E$_1$ (E$_1$~=~310--410~nm and
E$_2$~=~500--610~nm), is shown in the lower panel; the ordinate is 
the integrated photon content of each of the summed energy channels
(arbitrary scale).
}
\label{fig:ratio}
\end{center}
\end{figure}

\section{Results and Discussion}
\label{sec:results}

Fig.~\ref{fig:lightcurve} shows the light curve from the 50~min of 
data obtained on 6~Feb, acquired with a time resolution of 5~$\mu$s,
and folded into 128~phase bins ($\sim250~\mu$s per bin), 
with an arbitrary origin of zero phase, and without background 
subtraction (impractical due to the combination of poor seeing, small 
field, and fraction of unstable junctions; the overall system response 
is undetermined for similar reasons, combined with uncalibrated losses 
at the derotator entrance aperture). Examination of the light curve,
including the peaks, at finer temporal resolution down to about
30~$\mu$s per phase bin, reveals no significant sub-structure 
persisting over the observation interval.

Fig.~\ref{fig:ratio} shows the data divided into two separated
energy channels, corresponding to E$_1$~=~310--410~nm and
E$_2$~=~500--610~nm (energy channels 77--99 and 50--61 respectively). 
The profiles are normalised to the same relative 
intensities in the peaks, and are displaced vertically. For this choice of
energy bins, the ratio of photons is roughly E$_2$:E$_1$~=~4:1. The
resulting colour ratio, constructed as E$_2$--E$_1$/E$_2$+E$_1$ and
folded at the pulsar period, is shown in the lower part of the 
figure.

Wavelength-dependent variations across the peaks has been noted by
\cite*{efr+96}, based on observations spanning the UV to the infrared
K~band, and by \cite{srs98}. Not suprisingly given the form of these
reported variations, no significant variations with pulsar phase are
evident in our data. From the nebular emission-line spectrum reported
by \cite*{dav79} the [O\,{\sc ii}] doublet ($\lambda$ 3726, 3729) lies
within our extracted blue passband, while the red passband selected
comprises none of the strong emission lines in the red part of the
spectrum ([N\,{\sc ii}] 6584, H$\alpha$ 6563, [N\,{\sc ii}] 6584,
[Si\,{\sc ii}] 6717, 6731), and only marginally collects photons from
the [O\,{\sc iii}] doublet (4959, 5007). Our {\it a posteriori\/}
choice of energy channels for this ratio provides sufficient
wavelength separation to avoid `contamination' of each bin given the
low energy resolution of the device. Further discrimination of the
nebular contributions is limited by our presently modest energy
resolution, which is also insufficient to confirm the reality of the
broad absorption feature near 5920~\AA, so far noted only by \cite*{nmc+96}.

These observational results can be compared with predictions from models in 
which the observed flux in a given phase interval is a combination of 
emission from effectively disjointed physical regions, in which the 
sections contributing to the emission in a given phase interval depend 
on the viewing geometry (e.g.\ \cite{chr86a}, \cite{chr86b}). This 
mixing of physical regions acts to average the total emission,
and predicts that observed properties such as the energy ratio 
remain constant, or change modestly but rapidly close to the 
emission peak (\cite{rom96}). Higher S/N STJ data will be 
required to probe the small, rapid spectral index variations across 
the pulse peaks reported by \cite*{srs98}.

As the first application in astronomy of a superconducting tunnel
junction detector capable of providing intrinsic wavelength resolution
in the optical, the results are a modest indication of the
technology's capabilities for the future. Significant improvements in
performances, and in particular in the wavelength resolution, are
expected from the use of lower critical temperature superconductors in
the future.

\begin{acknowledgements}
We acknowledge the contributions of other members of the Astrophysics
Division of the European Space Agency at ESTEC involved in the optical
STJ development effort, in particular J.~Verveer and S.~Andersson (who
also provided technical and system engineering support at the
telescope) and P.~Verhoeve for evaluation of device performance. We
acknowledge D.~Goldie, R.~Hart and D.~Glowacka of Oxford Instruments
Thin Film Group for the fabrication of the array. We are grateful for
the assignment of engineering time at the William Herschel Telescope
of the ING, and we acknowledge the excellent support given to the
instrument's commissioning, in particular by P.~Moore and C.R.~Benn.
The analysis made use of the up-to-date Jodrell Bank Crab pulsar
timing results, maintained on the www by A.G.~Lyne, R.S.~Pritchard and
M.E.~Roberts. We thank G.~Vacanti and A.~Hazell for software updates
allowing our data to be processed within the FTOOLS/XRONOS
environment. We thank the referee, R.W.~Romani, for helpful comments.

\end{acknowledgements}


\end{document}